\documentclass[aps,twocolumn,amsmath,amssymb,a4paper]{revtex4}

\usepackage[utf8]{inputenc}
\usepackage[english]{babel}
\usepackage{amsmath} 
\usepackage{amsfonts} 
\usepackage{amssymb}
\usepackage{graphicx} \usepackage{color}
\usepackage{xcolor}

\newcommand{\mr}{\mathbf{r}}

\newcommand{\bra}[1]{\ensuremath{\langle #1 \vert}}
\newcommand{\ket}[1]{\ensuremath{\vert #1 \rangle}}
\renewcommand{\H}{\ensuremath{\text{H}}}
\renewcommand{\b}[1]{\ensuremath{\mathbf{#1}}}
\newcommand{\KS}{\ensuremath{\text{KS}}}
\renewcommand{\l}{\ensuremath{\lambda}}
\newcommand{\md}{\ensuremath{\text{md}}}
\newcommand{\lr}{\ensuremath{\text{lr}}}
\newcommand{\sr}{\ensuremath{\text{sr}}} \newcommand{\w}{w_{ee}}
\DeclareMathOperator{\erf}{erf} \DeclareMathOperator{\erfc}{erfc}
\newcommand{\Hxc}{\text{Hxc}}
\newcommand{\ee}{\text{ee}}
\newcommand{\x}{\text{x}}
\newcommand{\tc}{\text{c}}
\renewcommand{\S}{\text{S}}
\renewcommand{\P}{\text{P}}
\begin{document}

\title{Excited states from range-separated density-functional
  perturbation theory} 

\author{Elisa Rebolini$^{1,2,4}$}\email{erebolini@kjemi.uio.no}
\author{Julien Toulouse$^{1,2}$}
\author{Andrew M. Teale$^{3}$}
\author{Trygve Helgaker$^{4}$}
\author{Andreas Savin$^{1,2}$}
\affiliation{
$^1$Sorbonne Universit\'es, UPMC Univ Paris 06, UMR 7616, Laboratoire de Chimie Th\'eorique, F-75005 Paris, France\\
$^2$CNRS, UMR 7616, Laboratoire de Chimie Th\'eorique, F-75005 Paris, France\\
$^3$School of Chemistry, University of Nottingham, University Park, Nottingham NG7 2RD, United Kingdom\\
$^4$Centre for Theoretical and Computational Chemistry, Department of Chemistry, University of Oslo, P.O. Box 1033 Blindern, N-0315 Oslo, Norway}

\date{December 10, 2014}

\begin{abstract}
  We explore the possibility of calculating electronic excited states
  by using perturbation theory along a range-separated adiabatic connection.
  Starting from the energies of a partially interacting Hamiltonian,
  a first-order correction is defined with two variants of
  perturbation theory: a straightforward perturbation theory, and an
  extension of the G\"orling--Levy one that has the advantage of
  keeping the ground-state density constant at each order in the
  perturbation. Only the first, simpler, variant is tested here on the
  helium and beryllium atoms and on the dihydrogene molecule.
  The first-order correction within this perturbation theory
  improves significantly the total ground- and excited-state energies of the different
  systems. However, the excitation energies are mostly deteriorated
  with respect to the zeroth-order ones, which may be explained by the
  fact that the ionization energy is no longer correct for all
  interaction strengths. The second variant of the perturbation theory
  should improve these results but has not been tested yet along the
  range-separated adiabatic connection.
%  \\Keywords: Excitation energies; Range separation; Perturbation
%  theory; Adiabatic connection
\end{abstract}

\maketitle

%=====================================================================
% Introduction
%=====================================================================

\section{Introduction}

In density-functional theory (DFT) of quantum electronic systems, the
most widely used approach for calculating excitation energies is
nowadays linear-response time-dependent density-functional theory
(TDDFT) (see, e.g., Refs.~\onlinecite{Casida2009,
  Casida2012}). However, in spite of many successes, when applied with
the usual adiabatic semilocal approximations, linear-response TDDFT
has serious limitations for describing systems with static (or strong)
correlation~\cite{Gritsenko2000}, double or multiple
excitations~\cite{Maitra2004}, and Rydberg and charge-transfer
excitations~\cite{Casida1998,Dreuw2003}. Besides, the Hohenberg--Kohn
theorem~\cite{Hohenberg1964} states that the time-independent
ground-state density contains all the information about the system so
that time dependence is in principle not required to describe excited
states.

Several time-independent DFT approaches for calculating excitation
energies exist and are still being developed.  A first strategy
consists in simultaneously optimizing an ensemble of states. Such an
ensemble DFT was pioneered by Theophilou~\cite{Theophilou1979} and by
Gross, Oliveira and Kohn~\cite{Gross1988a} and is still a subject of
research~\cite{PasGidPer-PRA-13,Franck2013,Yang2014,Pribram-Jones2014},
but it is hampered by the absence of appropriate approximate ensemble
functionals. A second strategy consists in self-consistently
optimizing a single excited state. This approach was started by
Gunnarsson and Lundqvist~\cite{Gunnarsson1976}, who extended
ground-state DFT to the lowest-energy state in each symmetry class,
and developed into the pragmatic (multiplet-sum) $\Delta$SCF
method~\cite{Ziegler1977,Barth1979} (still in use
today~\cite{KowYosVoo-JCP-11}) and related
methods~\cite{FerAss-JCP-02, KryZie-JCTC-13, EvaShuTul-JPCA-13}. Great
efforts have been made by Nagy, G\"orling, Levy, Ayers and others to
formulate a rigorous self-consistent DFT of an arbitrary individual
excited state~\cite{Nag-IJQC-98, Gorling1999, Levy1999, Gor-PRL-00,
  NagLev-PRA-01, Nag-IJQC-04, Harbola2004, VitDelGor-JCP-05,
  Gorling2005, SamHar-JPB-06, GluLev-JCP-07, AyeLev-PRA-09,
  AyeLevNag-PRA-12} but a major difficulty is the need to develop
approximate functionals for a specific excited state (see
Ref.~\onlinecite{Harbola2012} for a proposal of such excited-state
functionals). A third strategy, first proposed by Grimme, consists in
using configuration-interaction (CI) schemes in which modified
Hamiltonian matrix elements include information from
DFT~\cite{Gri-CPL-96,GriWal-JCP-99,BecStaBurBla-CP-08,KadVoo-JCP-10}.

Finally, a fourth possible approach, proposed by
G\"orling~\cite{Gorling1996}, is to calculate the excitation energies
from G\"orling--Levy (GL) perturbation theory~\cite{GorLev-PRB-93,
  Gorling1995} along the adiabatic connection using the
non-interacting Kohn--Sham (KS) Hamiltonian as the zeroth-order
Hamiltonian. In this approach, the zeroth-order approximation to the
exact excitation energies is provided by KS orbital energy differences
(which, for accurate KS potentials, is known to be already a fairly
good approximation~\cite{Savin1998}). It can be improved upon by
perturbation theory at a given order in the coupling constant of the
adiabatic connection. Filippi, Umrigar, and Gonze~\cite{Filippi1997}
showed that the GL first-order corrections provide a factor of two
improvement to the KS zeroth-order excitation energies for the He,
Li$^+$, and Be atoms when using accurate KS potentials. For (nearly)
degenerate states, Zhang and Burke~\cite{Zhang2004} proposed to use
degenerate second-order GL perturbation theory, showing that it works
well on a simple one-dimensional model. This approach is conceptually
simple as it uses the standard adiabatic connection along which the
ground-state density is kept constant (in contrast to approaches
introducing generalized adiabatic connections keeping an excited-state
density constant~\cite{Nag-IJQC-98, Gorling1999, Gor-PRL-00,
  Gorling2005}). In spite of promising early results, this approach
has not been pursued further, perhaps because it can be considered an
approximation to TDDFT~\cite{Gonze1999}.

In this work, we explore further this density-functional
perturbation-theory approach for calculating excitation energies,
introducing one key modification in comparison to the earlier work of
Refs.~\onlinecite{Gorling1996,Filippi1997}: As a zeroth-order
Hamiltonian, instead of using the non-interacting KS Hamiltonian, we
use a {\it partially interacting Hamiltonian} incorporating the {\it
  long-range} part of the Coulomb electron--electron interaction,
corresponding to an intermediate point along a range-separated
adiabatic connection~\cite{Savin1996,Yan-JCP-98,Pollet2003,Savin2003,
  Toulouse2004,Rebolini2014}. The partially interacting zeroth-order
Hamiltonian is of course closer to the exact Hamiltonian than is the
non-interacting KS Hamiltonian, thereby putting less demand on the
perturbation theory. In fact, the zeroth-order Hamiltonian can already
incorporate some static correlation.

The downside is that a many-body method such as CI theory is required
to generate the eigenstates and eigenvalues of the zeroth-order
Hamiltonian. However, if the partial electron--electron interaction is
only a relatively weak long-range interaction, we would expect a
faster convergence of the eigenstates and eigenvalues with respect to
the one- and many--electron CI expansion than for the full Coulomb
interaction~\cite{Toulouse2004,FraMusLupTou-JJJ-XX}, so that a small CI or
multi-configuration self-consistent field (MCSCF) calculation would be
sufficiently accurate.

When using a semi-local density-functional approximation for the
effective potential of the range-separated adiabatic connection, the
presence of an explicit long-range electron--electron interaction in
the zeroth-order Hamiltonian also has the advantage of preventing the
collapse of the high-lying Rydberg excitation energies. In contrast to
adiabatic TDDFT, double or multiple excitations can be described with
this density-functional perturbation-theory approach, although this
possibility was not explored in Refs.~\onlinecite{Gorling1996,
  Filippi1997}. Finally, approximate excited-state wave functions are
obtained, which is useful for interpretative analysis and for the
calculation of properties.

We envisage using this density-functional perturbation theory to
calculate excited states after a range-separated ground-state
calculation combining a long-range CI~\cite{Leininger1997,Pollet2002}
or long-range MCSCF~\cite{Fromager2007,FroReaWahWahJen-JCP-09}
treatment with a short-range density functional. This would be a
simpler alternative to linear-response range-separated MCSCF
theory~\cite{Fromager2013, HedHeiKneFroJen-JCP-13} for calculations of
excitation energies. In this work, we study in detail the two variants
of range-separated density-functional perturbation theory and test the
first, simpler variant on the He and Be atoms and the H$_2$ molecule,
using accurate calculations along a range-separated adiabatic
connection without introducing density-functional approximations.

Both variants of the range-separated perturbation 
theory are presented in Section~\ref{sec:perturbation theory}.  
Except for the finite basis approximation, no other approximation is introduced and the
computational details can be found in Section~\ref{sec:computational}.
Finally, the results obtained for the He and Be atoms, and
for the H$_2$ molecule are discussed in Section~\ref{sec:results}.
Section~\ref{sec:conclusion} contains our conclusions.

%=====================================================================
% Theory
%=====================================================================
 \section{Range-separated density-functional perturbation theory}
\label{sec:perturbation theory}
\subsection{Range-separated ground-state density-functional theory}
\label{sec:rsdft}

In range-separated DFT (see, e.g., Ref.~\onlinecite{Toulouse2004}),
the exact ground-state energy of an $N$-electron system is obtained by
the following minimization over normalized multi-determinantal wave
functions $\Psi$
\begin{eqnarray}
  E_0 &=& \min_{\Psi} \Bigl\{ \bra{\Psi} \hat{T} + \hat{V}_\text{ne} +
  \hat{W}_{\ee}^{\lr,\mu} \ket{\Psi} +
  \bar{E}_{\Hxc}^{\sr,\mu}[n_{\Psi}]\Bigl\}, \nonumber\\
\label{EminPsi}
\end{eqnarray}
where we have introduced the kinetic-energy operator $\hat{T}$, the
nuclear.attraction operator $\hat{V}_\text{ne} = \int
v_\text{ne}(\b{r}) \hat{n}(\b{r}) \mathrm d\b{r}$ written in terms of the
density operator $\hat{n}(\b{r})$, a long-range (lr)
electron--electron interaction
\begin{eqnarray}
  \hat{W}_{\ee}^{\lr,\mu} \!=\! \frac{1}{2} \iint
  w_{\ee}^{\lr,\mu}(r_{12}) \hat{n}_2(\b{r}_1,\b{r}_2) \mathrm d\b{r}_1
   \mathrm d\b{r}_2,
\end{eqnarray}
written in terms of the error-function interaction
$w_{\ee}^{\lr,\mu}(r)\!  =\!\erf(\mu r)/r$ and the pair-density
operator $\hat{n}_2(\b{r}_1,\b{r}_2)$, and finally the corresponding
complement short-range (sr) Hartree--exchange--correlation density
functional $\bar{E}_{\Hxc}^{\sr,\mu}[n_\Psi]$ evaluated at the density
of $\Psi$: $n_\Psi(\b{r}) = \bra{\Psi} \hat{n}(\b{r}) \ket{\Psi}$. The
parameter $\mu$ in the error function controls the separation range,
with $1/\mu$ acting as a smooth cut-off radius.

The Euler--Lagrange equation for the minimization of
Eq.~(\ref{EminPsi}) leads to the (self-consistent) eigenvalue equation
\begin{equation}
  \hat{H}^{\lr,\mu} | \Psi_0^{\mu} \rangle = {\cal E}_0^{\mu} |
  \Psi_0^{\mu} \rangle,
\end{equation}
where $\Psi_0^{\mu}$ and ${\cal E}_0^{\mu}$ are taken as the
ground-state wave function and associated energy of the partially
interacting Hamiltonian (with an explicit long-range
electron--electron interaction)
\begin{equation}
  \hat{H}^{\lr,\mu} = \hat{T} + \hat{V}_\text{ne} +
  \hat{W}^{\lr,\mu}_{\ee} + \hat{\bar{V}}_{\Hxc}^{\sr,\mu},
  \label{Hmu}
\end{equation}
which contains the short-range Hartree--exchange--correlation potential operator,
\begin{equation}
  \hat{\bar{V}}_{\H xc}^{\sr,\mu}= \int \bar{v}^{\sr,\mu}_{\H
    xc}[n_0](\mr) \hat{n}(\mr) \mathrm d\mr,
\end{equation}
where $\bar{v}^{\sr,\mu}_{\H xc}[n](\mr) = \delta
\bar{E}_{\Hxc}^{\sr,\mu}[n] / \delta n(\mr)$, evaluated at the
ground-state density of the physical system $n_0(\b{r}) =
\bra{\Psi_0^{\mu}} \hat{n}(\b{r}) \ket{\Psi_0^{\mu}}$ for all $\mu$.

For $\mu=0$, the Hamiltonian $\hat{H}^{\lr,\mu}$ of Eq.~(\ref{Hmu})
reduces to the standard non-interacting KS Hamiltonian,
$\hat{H}^{\lr,\mu=0}=\hat{H}^{\KS}$, whereas, for $\mu\to\infty$, it
reduces to the physical Hamiltonian $\hat{H}^{\lr,\mu \to
  \infty}=\hat{H}$. Therefore, when varying the parameter $\mu$
between these two limits, the Hamiltonian $\hat{H}^{\lr,\mu}$ defines
a range-separated adiabatic connection, linking the non-interacting KS
system to the physical system with the ground-state density kept
constant (assuming that the exact short-range
Hartree--exchange--correlation potential
$\bar{v}_{\Hxc}^{\sr,\mu}(\mr)$ is used).

\subsection{Excited states from perturbation theory}

Excitation energies in range-separated DFT can be obtained by
linear-response theory starting from the (adiabatic) time-dependent
generalization of Eq.\;(\ref{EminPsi})~\cite{FroKneJen-JCP-13}.  Here,
the excited states and their associated energies are instead obtained
from time-independent many-body perturbation theory.  In standard KS
theory, the single-determinant eigenstates and associated energies of
the non-interacting KS Hamiltonian,
\begin{equation}
  \hat{H}^{\KS} | \Phi_k^{\KS} \rangle = {\cal E}_k^{\KS} |
  \Phi_k^{\KS} \rangle,
\end{equation}
where $\hat{H}^{\KS} = \hat{T} + \hat{V}_\text{ne} + \hat{V}_{\Hxc}$,
give a first approximation to the eigenstates and associated energies
of the physical Hamiltonian. To calculate excitation energies, two
variants of perturbation theory using the KS Hamiltonian as
zeroth-order Hamiltonian have been proposed~\cite{Gorling1996,
  Filippi1997}. We here extend these two variants of perturbation
theory to range-separated DFT. As a first approximation, it is natural
to use the excited-state wave functions and energies of the long-range
interacting Hamiltonian
\begin{equation}
  \hat{H}^{\lr,\mu} | \Psi_k^{\mu} \rangle = {\cal E}_k^{\mu} |
  \Psi_k^{\mu} \rangle,
\label{zerothordereq}
\end{equation}
where $\hat{H}^{\lr,\mu} = \hat{T} + \hat{V}_\text{ne} +
\hat{\bar{V}}^{\sr,\mu}_{\Hxc} + \hat{W}_{\ee}^{\lr,\mu}$ is the same
Hamiltonian that appears in Eq.~(\ref{Hmu}) with the short-range
Hartree--exchange--correlation potential
$\hat{\bar{V}}^{\sr,\mu}_{\Hxc} = \int
\bar{v}_{\Hxc}^{\sr,\mu}[n_0]\hat{n}(\mr) \mathrm d\mr$ evaluated at the
ground-state density $n_0$. These excited-state wave functions and
energies can then be improved upon by defining perturbation theories
in which the Hamiltonian $\hat{H}^{\lr,\mu}$ is used as the
zeroth-order Hamiltonian.

\subsubsection{First variant of perturbation theory}

The simplest way of defining such a perturbation theory is to
introduce the following Hamiltonian dependent on the coupling constant
$\l$
\begin{equation}
  \hat{H}^{\mu,\l} = \hat{H}^{\lr,\mu} + \l \hat{W}^{\sr,\mu},
  \label{Hlmu}
\end{equation}
where the short-range perturbation operator is
\begin{equation}
  \hat{W}^{\sr,\mu} = \hat{W}_{\ee}^{\sr,\mu} -
  \hat{\bar{V}}_{\Hxc}^{\sr,\mu},
\end{equation}
with the short-range electron--electron interaction
$\hat{W}_{\ee}^{\sr,\mu} = (1/2) \iint w_{\ee}^{\sr,\mu}(r_{12})
\hat{n}_2(\mr_1,\mr_2) \mathrm d\mr_1 \mathrm d\mr_2$ defined with the complementary
error-function interaction $w_{\ee}^{\sr,\mu}(r)\! =\!\erfc(\mu
r)/r$. When varying $\l$, Eq.~(\ref{Hlmu}) sets up an adiabatic
connection linking the long-range interacting Hamiltonian at $\l=0$,
$\hat{H}^{\mu,\l=0} = \hat{H}^{\lr,\mu}$, to the physical Hamiltonian
at $\l=1$, $\hat{H}^{\mu,\l=1} = \hat{H}$, for all $\mu$. However, the
ground-state density is {\it not kept constant} along this adiabatic
connection.

The exact eigenstates and associated eigenvalues of the physical
Hamiltonian can be obtained by standard Rayleigh--Schr\"odinger
perturbation theory---that is, by Taylor expanding the eigenstates and
eigenvalues of the Hamiltonian $\hat{H}^{\mu,\l}$ in $\lambda$ and
setting $\l=1$:
\begin{subequations}
  \begin{align}
    \ket{\Psi_k} & = \ket{\Psi_k^{\mu}} + \sum_{n=1}^{\infty}
    \ket{\Psi_k^{\mu,(n)}}, \\ E_k & = {\cal E}_k^{\mu} +
    \sum_{n=1}^{\infty} E_k^{\mu,(n)},
     \label{eq:TaylorPsi}
  \end{align}
\end{subequations}
where $\Psi_k^{\mu} \equiv \Psi_k^{\mu,(0)}$ and ${\cal E}_k^{\mu}
\equiv E_k^{\mu,(0)}$ act as zeroth-order eigenstates and
energies. Using orthonormalized zeroth-order eigenstates $\langle
\Psi_k^{\mu} | \Psi_l^{\mu} \rangle = \delta_{kl}$ and assuming
non-degenerate zeroth-order eigenstates, the first-order energy
correction for the state $k$ becomes
\begin{equation}
  E_k^{\mu,(1)} = \langle \Psi_k^{\mu} | \hat{W}^{\sr,\mu}
  |\Psi_k^{\mu} \rangle.
 \label{eq:E1}
\end{equation}
As usual, the zeroth-plus-first-order energy is simply the expectation
value of the physical Hamiltonian over the zeroth-order eigenstate:
\begin{equation}
  E_k^{\mu,(0+1)}= {\cal E}_k^{\mu} + E_k^{\mu,(1)} = \langle
  \Psi_k^{\mu}| \hat{H} |\Psi_k^{\mu} \rangle.
\label{eq:E0+1}
\end{equation}
This expression is a multi-determinantal extension of the
exact-exchange KS energy expression for the state $k$, proposed and
studied for the ground state in Refs.~\onlinecite{TouGorSav-TCA-05,
  GorSav-IJQC-09a, Stoyanova2013}. The second-order energy correction
is given by
\begin{equation}
  E_k^{\mu,(2)} = - \sum_{l \neq k} \frac{|\bra{\Psi_l^{\mu}}
    \hat{W}^{\sr,\mu} \ket{\Psi_k^{\mu}}|^2}{{\cal E}_l^{\mu} - {\cal
      E}_k^{\mu}},
\label{Eimu2}
\end{equation}
whereas the first-order wave-function correction is given by (using
intermediate normalization so that $\langle \Psi_k^{\mu} |
\Psi_k^{\mu,(n)} \rangle = 0$ for all $n\geq 1$)
\begin{equation}
  \ket{\Psi_k^{\mu,(1)}} = - \sum_{l \neq k} \frac{\bra{\Psi_l^{\mu}}
    \hat{W}^{\sr,\mu} \ket{\Psi_k^{\mu}}}{{\cal E}_l^{\mu} - {\cal
      E}_k^{\mu}} \ket{\Psi_l^{\mu}}.
  \label{Psimu1}
\end{equation}
For $\mu=0$, this perturbation theory reduces to the first variant of
the KS perturbation theory studied by Filippi {\it et al.}, see
Eq.~(5) of Ref.~\onlinecite{Filippi1997}.

%=====================================================================
We now give the behaviors of the zeroth+first-order energies with respect $\mu$ near the KS system ($\mu=0$) and near the physical system ($\mu\to\infty$), which are useful to understand the numerical results in Section~\ref{sec:results}.
The total energies up to the first order of the perturbation theory
are given by the expectation value of the full Hamiltonian over the
zeroth-order wave functions in Eq~\eqref{eq:E1}.  Using the Taylor
expansion of the wave function $\Psi_k^{\mu} = \Phi_k^{\KS} + \mu^3
\Psi_k^{(3)} +\mathcal{O}(\mu^5)$ around the KS wave
function~\cite{Rebolini2014}, it implies that the zeroth+first-order
energies are thus given by
\begin{equation}
  \begin{split}
    E_k^{\mu,(0+1)}  =
    & \,\langle \Phi_k^{\KS} | \hat{H} | \Phi_k^{\KS} \rangle 
     \\ &+ 2\mu^3 
      \langle \Phi_k^{\KS} | \hat{H} | \Psi_k^{(3)} \rangle 
    + \mathcal{O}(\mu^5),
    \label{eq:E0+1 0}
  \end{split}
\end{equation}
where $ \Psi_k^{(3)}$ is the contribution entering at the third
power of $\mu$ in the zeroth-order wave function.  

>From the asymptotic expansion of the wave function $\Psi_k^\mu = \Psi_k +
\mu^{-2}\Psi_k^{(-2)} + \mathcal{O}(\mu^{-3})$, which is valid almost
everywhere when $\mu\to\infty$ (the electron-electron coalescence needs to be treated carefully)
~\cite{Rebolini2014}, the first correction to the
zeroth+first-order energies enter at the fourth power of $\mu$
\begin{equation}
  E_k^{\mu,(0+1)} = E_k + \dfrac{1}{\mu^4}E_k^{(0+1,-4)} +
  \mathcal{O}\left(\dfrac{1}{\mu^6} \right),
  \label{eq:E0+1infty}
\end{equation}
where $E_k^{(0+1,-4)}$ is the contribution entering at the fourth power
of $1/\mu$.

\subsubsection{Second variant of perturbation theory}
\label{sec:secondvariant}

A second possibility is to define a perturbation theory based on a
slightly more complicated adiabatic connection, in which the
ground-state density is {\it kept constant} between the long-range
interacting Hamiltonian and the physical Hamiltonian, see
Appendix~\ref{app:doubleadia}. The Hamiltonian of Eq.~(\ref{Hlmu}) is
then replaced by
\begin{equation}
  \hat{H}^{\mu,\l} = \hat{H}^{\lr,\mu} + \l \hat{W}^{\sr,\mu} -
  \hat{V}_{\tc,\md}^{\sr,\mu,\l},
\label{Hlmu2}
\end{equation}
where $\hat{W}^{\sr,\mu}$ is now defined as
\begin{equation}
    \hat{W}^{\sr,\mu}  = \hat{W}_{\ee}^{\sr,\mu} - \hat{V}_{\H \x,\md}^{\sr,\mu},
\label{Wsrmu2}
\end{equation}
in terms of a short-range ``multi-determinantal (md) Hartree--exchange''
potential operator 
\begin{equation}
  \hat{V}_{\H \x,\md}^{\sr,\mu}= \int \dfrac{\delta E_{\H
  \x,\md}^{\sr,\mu}[n_0]}{\delta n(\mr)} \hat{n}(\mr) \,\mathrm d\b{r},
\end{equation}
and a short-range ``multi-determinantal correlation''
potential operator 
\begin{equation}
  \hat{V}_{\tc,\md}^{\sr,\mu,\l}= \int \dfrac{\delta
  E_{\tc,\md}^{\sr,\mu,\l}[n_0]}{\delta n(\mr)} \hat{n}(\mr) \,\mathrm d\b{r},
\end{equation}
that depends {\it non-linearly} on $\l$ so that the ground-state
density $n_0$ is kept constant for all $\mu$ and $\l$. The density
functionals $E_{\H \x,\md}^{\sr,\mu}[n]$ and
$E_{\tc,\md}^{\sr,\mu,\l}[n]$ are defined in
Appendix~\ref{app:doubleadia}.

One can show that, for non-degenerate ground-state wave functions
$\Psi_0^{\mu}$, the expansion of $\hat{V}_{\tc,\md}^{\sr,\mu,\l}$ in
$\l$ for $\l\to 0$ starts at second order:
\begin{equation}
  \hat{V}_{\tc,\md}^{\sr,\mu,\l} = \l^2 \,
  \hat{V}_{\tc,\md}^{\sr,\mu,(2)} + \cdots.
\end{equation}
 Hence, the Hamiltonian of Eq.~(\ref{Hlmu2}) properly reduces to the
 long-range Hamiltonian at $\l=0$, $\hat{H}^{\mu,\l=0} =
 \hat{H}^{\lr,\mu}$, whereas, at $\l=1$, it correctly reduces to the
 physical Hamiltonian, $\hat{H}^{\mu,\l=1} = \hat{H}$. This is so
 because the short-range Hartree--exchange--correlation potential in
 the Hamiltonian $\hat{H}^{\lr,\mu}$ can be decomposed as
\begin{equation}
  \hat{\bar{V}}_{\Hxc}^{\sr,\mu} = \hat{V}_{\H \x,\md}^{\sr,\mu} +
  \hat{\bar{V}}_{\tc,\md}^{\sr,\mu},
\label{Vhxcsrmudecompmd}
\end{equation}
where $\hat{\bar{V}}_{\tc,\md}^{\sr,\mu} =
\hat{V}_{\tc,\md}^{\sr,\mu,\l=1}$ is canceled by the perturbation
terms for $\l=1$. Equation~(\ref{Vhxcsrmudecompmd}) corresponds to an
alternative decomposition of the short-range
Hartree--exchange--correlation energy into ``Hartree--exchange'' and
``correlation'' contributions based on the multi-determinantal wave
function $\Psi_0^\mu$ instead of the single-determinant KS wave
function
$\Phi_0^\KS$~\cite{TouGorSav-TCA-05,GorSav-IJQC-09a,Stoyanova2013},
which is more natural in range-separated DFT.  This decomposition is
especially relevant here since it separates the perturbation into a
``Hartree--exchange'' contribution that is linear in $\l$ and a
``correlation'' contribution containing all the higher-order terms in
$\l$.

The first-order energy correction is still given by
Eq.~(\ref{eq:E1}) but with the perturbation operator of
Eq.~(\ref{Wsrmu2}), yielding the following energy up to first order:
\begin{equation}
  E_k^{\mu,(0+1)}= {\cal E}_k^{\mu} + E_k^{\mu,(1)} = \langle
  \Psi_k^{\mu}| \hat{H} + \hat{\bar{V}}_{\tc,\md}^{\sr,\mu}
  |\Psi_k^{\mu} \rangle.
\end{equation}
The second-order energy correction of Eq.~(\ref{Eimu2}) becomes 
\begin{equation}
  \begin{split}
    E_k^{\mu,(2)} = & \, - \sum_{l \neq k} \frac{|\bra{\Psi_l^{\mu}}
      \hat{W}^{\sr,\mu} \ket{\Psi_k^{\mu}}|^2}{{\cal E}_l^{\mu} - {\cal
        E}_k^{\mu}} \\
    & - \langle \Psi_k^{\mu}|
    \hat{V}_{\tc,\md}^{\sr,\mu,(2)} |\Psi_k^{\mu} \rangle,
    \end{split}
\end{equation}
whereas the expression of the first-order wave function correction is
still given by Eq.~(\ref{Psimu1}) but with the perturbation operator of
Eq.~(\ref{Wsrmu2}).

For $\mu=0$, this density-fixed perturbation theory reduces to the
second variant of the KS perturbation theory proposed by
G\"orling~\cite{Gorling1996} and studied by Filippi {\it et
  al.}~[Eq.~(6) of Ref.~\onlinecite{Filippi1997}], which is simply the
application of GL perturbation theory~\cite{GorLev-PRB-93,Gorling1995}
to excited states. In Ref.~\onlinecite{Filippi1997}, it was found that
the first-order energy corrections in density-fixed KS perturbation
theory provided on average a factor of two improvement on the KS
zeroth-order excitation energies for the He, Li$^+$, and Be atoms when
using accurate KS potentials. By contrast, the first-order energy
corrections in the first variant of KS perturbation theory, without a
fixed density, deteriorated on average the KS excitation energies.

\begin{figure*}
  \centering
  %\hspace*{-2.5cm}
  \includegraphics[scale=0.9]{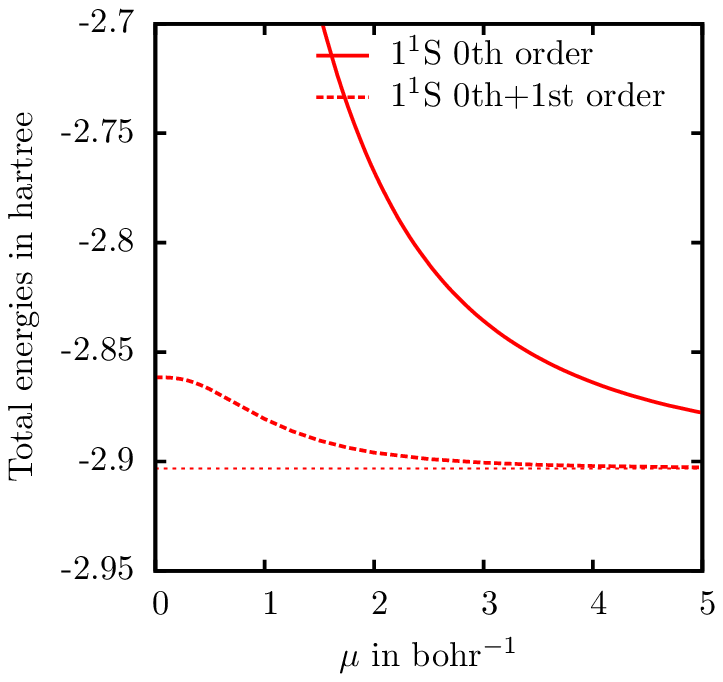} 
  %\hspace*{-2.5cm}
  \includegraphics[scale=0.9]{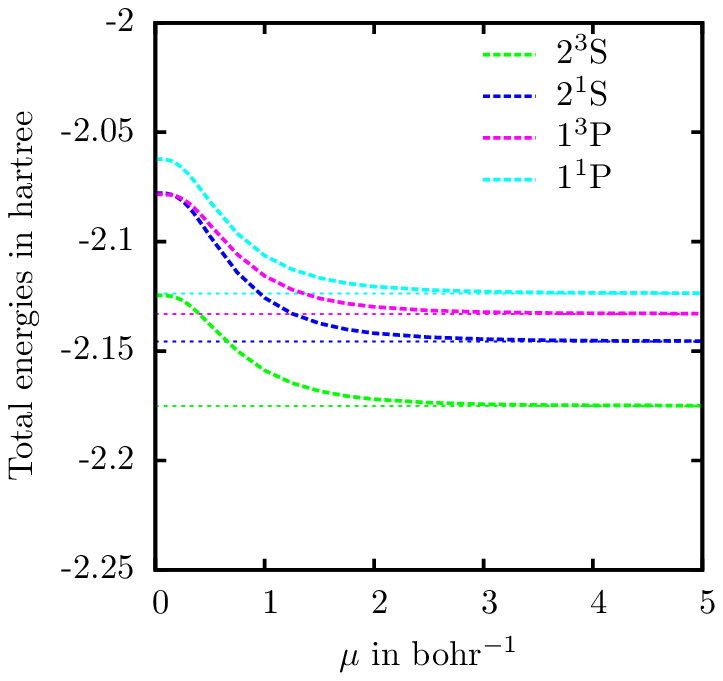}
  %\hspace*{-2.5cm}
  \caption[0th+1st order total energies of He] {Zeroth+first-order
    ground- (left) and excited-state (right) total energies ${
     E}_k^{\mu,(0+1)}$ (in hartree) of the helium atom as a function
    of $\mu$ (in bohr$^{-1}$). The zeroth-order energy ${\cal
      E}_0^{\mu}$ is recalled for the ground state in plain line and
    the total energies of the physical system $E_k$ are plotted as
    horizontal dotted lines.
    \label{fig:he_state_tacpv5z_svd07_0+1}
  }
\end{figure*}

The good results obtained with the second variant of KS perturbation
theory may be understood from that fact that, in GL perturbation
theory, the ionization potential remains exact to all orders in
$\l$. In fact, this nice feature of GL theory holds also with range
separation, so that the second variant of range-separated perturbation
theory should in principle be preferred. However, it requires the
separation of the short-range Hartree--exchange--correlation potential
into the ``multi-determinantal Hartree--exchange'' and
``multi-determinantal correlation'' contributions (according to
Eq.~(\ref{Vhxcsrmudecompmd})), which we have not done for accurate
potentials or calculations along the double adiabatic connection with
a partial interaction defined by $\hat{W}_{\ee}^{\lr,\mu} + \lambda
\hat{W}_{\ee}^{\sr,\mu}$ (cf.~Appendix~\ref{app:doubleadia}). We
therefore consider only the first variant of range-separated
perturbation theory here but note that the second variant can be
straightforwardly applied with density-functional
approximations---using, for example, the local-density approximation
that has been constructed for the ``multi-determinantal correlation''
functional~\cite{TouGorSav-TCA-05,Paziani2006}.

\begin{figure*}
  \centering
  %\hspace*{-2.5cm}
  \includegraphics[scale=0.9]{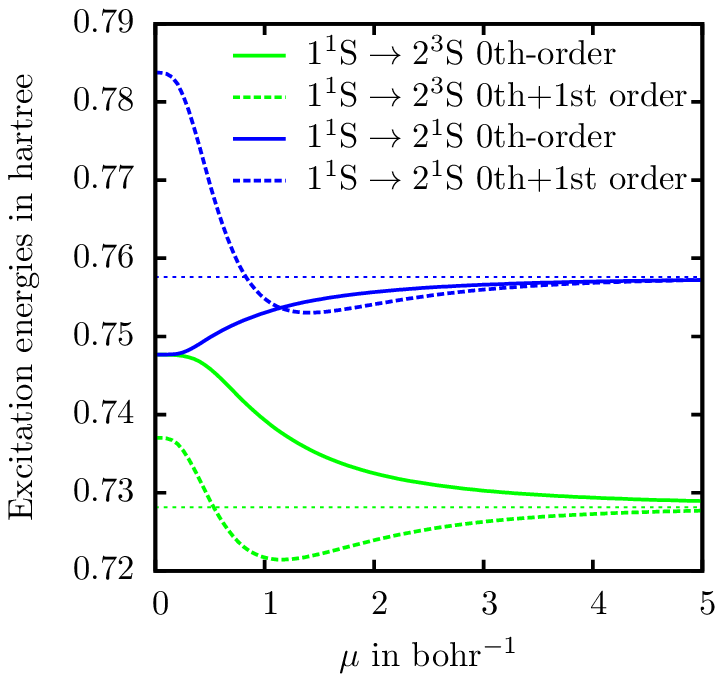}
  %\hspace*{-2.5cm}
  \includegraphics[scale=0.9]{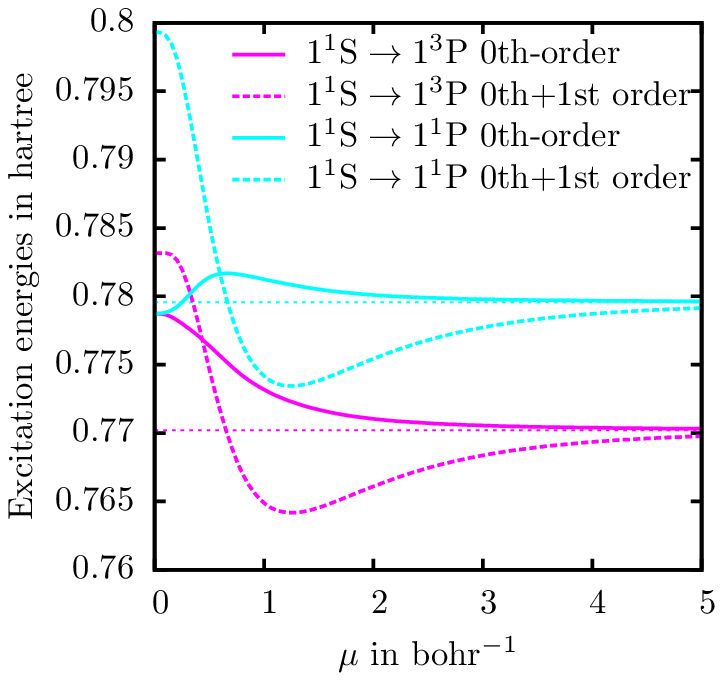}
  %\hspace*{-2.5cm}
  \caption[0th+1st order excitation energies of He]{Zeroth-order
    (plain line) excitation energies $\Delta {\cal E}_k^{\mu}={\cal E}_k^{\mu}-{\cal E}_0^{\mu}$ and
    zeroth+first-order (dashed line) excitation energies $\Delta {
      E}_k^{\mu,(0+1)}=E_k^{\mu,(0+1)} - E_0^{\mu,(0+1)}$ (in hartree) of the helium atom as a function
    of $\mu$ (in bohr$^{-1}$). The excitation energies of the physical
    system $\Delta E_k=E_k-E_0$ are plotted as horizontal dotted lines.
    \label{fig:he_exc_tacpv5z_svd07_0+1}
  }
\end{figure*}

\section{Computational details}
\label{sec:computational}

Calculations were performed for the He and Be atoms and the H$_2$
molecule with a development version of the DALTON
program~\cite{Dal-PROG-11}, see
Refs.~\onlinecite{TeaCorHel-JCP-09,TeaCorHel-JCP-10,TeaCorHel-JCP-10b}.
Following the same settings as in Ref.~\onlinecite{Rebolini2014}, 
a full CI (FCI) calculation was first carried out to get the exact
ground-state density within the basis set considered. Next, a Lieb
optimization of the short-range potential $v^{\sr,\mu}(\b{r})$ was
performed to reproduce the FCI density with the long-range
electron--electron interaction $\w^{\lr,\mu}(r_{12})$. Then, an FCI
calculation was done with the partially-interacting Hamiltonian
constructed from $\w^{\lr,\mu}(r_{12})$ and $v^{\sr,\mu}(\b{r})$ to
obtain the zeroth-order energies and wave functions according to Eq.~(\ref{zerothordereq}).
Finally, the zeroth+first order energies were calculated according to Eq.~(\ref{eq:E0+1}).
The basis sets used were: uncontracted t-aug-cc-pV5Z for He, uncontracted d-aug-cc-pVDZ for Be,
and uncontracted d-aug-cc-pVTZ for H$_2$.

%=====================================================================
% Results and discussions
%=====================================================================
\section{Results and discussion}
\label{sec:results}
%=====================================================================
\subsection{Helium atom}

The ground- and excited-state total energies to first order of 
helium along the range-separated adiabatic connection are shown
in Figure~\ref{fig:he_state_tacpv5z_svd07_0+1}. In the KS 
limit, when $\mu=0$, the total energies are significantly
improved with respect to the zeroth-order ones.
%given in Figure~\ref{fig:he_tacpv5z_state_svd07_0}.  
In fact, as shown for the ground-state energy, the zeroth-order total
energies were off by approximately 1.2 hartree with respect to the
energies of the physical system. When the first-order correction is
added, the error becomes smaller than 0.06 hartree for all states.
Moreover, the singlet and triplet excited-state energies are no longer
degenerate.  With increasing the range-separation parameter $\mu$, a
faster convergence towards the total energies of the physical system
is also observed for all states.

The description of the total energies is therefore much improved with
the addition of the first-order correction. The linear
term in $\mu$ present in the zeroth-order total energies~\cite{Rebolini2014} is no longer there 
for the zeroth+first order total energies, which instead depend on the
the third power of $\mu$ for small $\mu$ (cf. Eq.~\eqref{eq:E0+1 0}). At large $\mu$, the
error relative to the physical energies enters as $1/\mu^4$ rather
than as $1/\mu^2$ in the zeroth-order case, explaining the observed
faster convergence of the first-order energies.

\begin{figure}[b]
  \centering
  %\hspace*{-1.1cm}
  \includegraphics[scale=0.9]{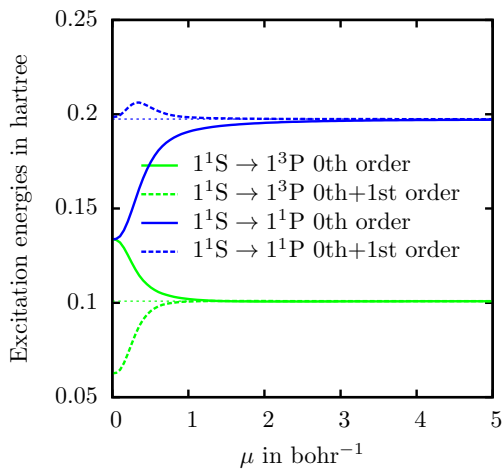}
  \caption[0th+1st order excitation energies of Be]{Valence excitation
    energies of the beryllium atom (in hartree) at zeroth order
    $\Delta {\cal E}_k^{\mu}$ (plain line) and zeroth+first order
    $\Delta {E}_k^{\mu,(0+1)}$ (dashed line), as a function of
    $\mu$ (in bohr$^{-1}$).  The excitation energies of the physical
    system $\Delta E_k$ are plotted as horizontal dotted lines.
    \label{fig:be_dacpvdz_exc_0+1}
  }
\end{figure}

\begin{figure*}
  \centering
  %\hspace*{-2.5cm}
  \includegraphics[scale=0.9]{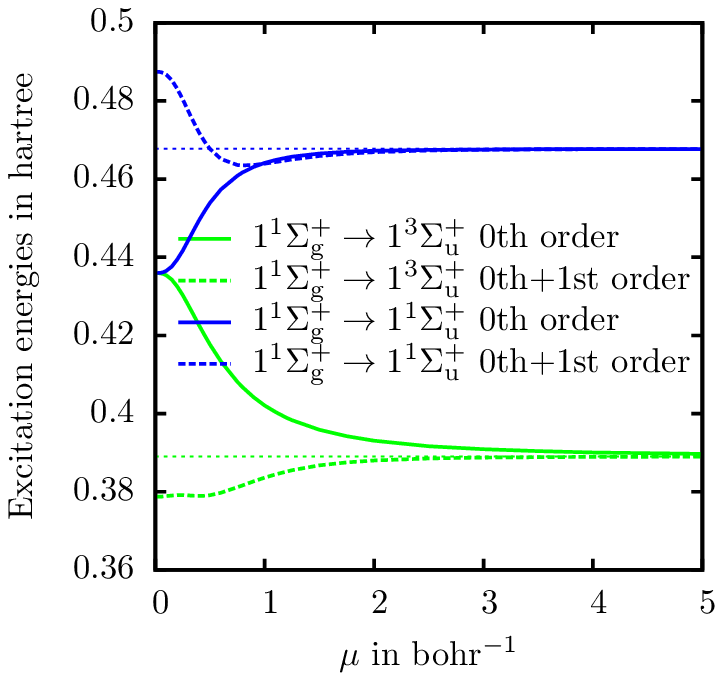}
  %\hspace*{-2.5cm}
  \includegraphics[scale=0.9]{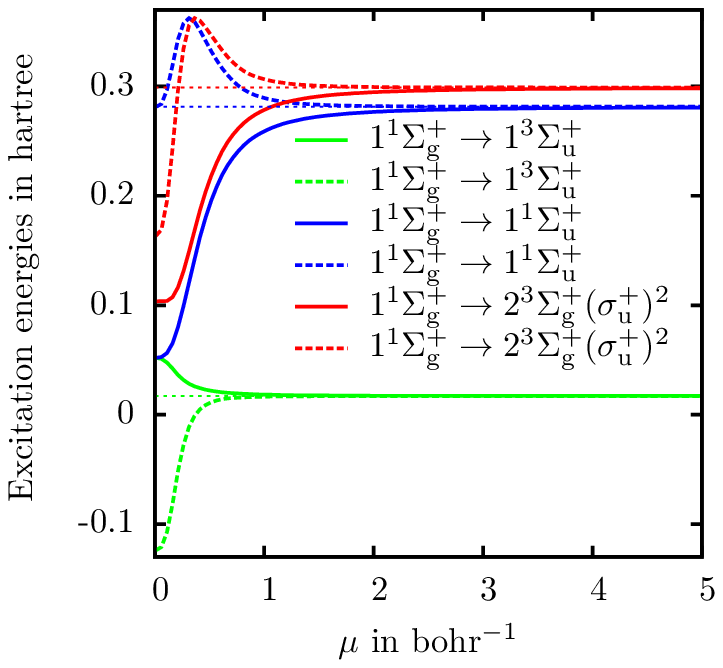}
  %\hspace*{-2.5cm}
  \caption[0th+1st order excitation energies of H$_2$ at
    $R_{eq}$]{Zeroth-order $\Delta {\cal E}_k^{\mu}$ (plain line) and
    zeroth+first-order $\Delta {E}_k^{\mu,(0+1)}$ (dashed line)
    excitation energies of the hydrogen molecule (in hartree) as a
    function of $\mu$ in bohr$^{-1}$ at the equilibrium distance (left)
    and three times the equilibrium distance (right). The excitation
    energies of the physical system $\Delta E_k$ are plotted as
    horizontal dotted lines.
    \label{fig:h2_eq_dacpvtz_exc_0+1}
  }
\end{figure*}

The excitation energies of the helium atom correct to zeroth and first 
orders are plotted in Figure~\ref{fig:he_exc_tacpv5z_svd07_0+1}.  
As previously noted, at $\mu=0$, the degeneracy of the zeroth-order singlet and triplet excitation energies 
is lifted by the first-order correction, 
However, the excitation energies correct to first order 
overestimate the physical excitation energies
by 0.1--0.2 hartree such that the error is actually
larger than at zeroth order. For the $1^1\S \to 1^3\P$ excitation
energy, the correction is even going in the wrong
direction and the singlet--triplet splitting is too large by about a factor 1.5.

When the very long-range part of the Coulombic interaction is switched
on with positive $\mu$ close to $0$, the initial overestimation is corrected. In fact, for small 
$\mu$, all the excitation energies decrease in the third power of
$\mu$ which is in agreement with Eq.~\eqref{eq:E0+1 0}. When
$\mu \simeq 0.5-1$, this correction becomes too large and
the excitation energies of the partially interacting system become
lower than their fully interacting limits. 
As $\mu$ increases further so that more
interaction is included, the excitation energies 
converge toward their fully interacting values from below.  
The zeroth-order excitation energies, which do not
oscillate for small $\mu$, converge monotonically toward
their physical limit and are on average more accurate than the
zeroth+first order excitation energies. In short, the first-order
correction does not improve excitation
energies, although total energies are improved.

The inability of the first-order correction to improve excitation
energies should be connected to the fact that, since the
ground-state density is not kept constant at each order in the
perturbation, the ionization potential is no longer constant to first order
along the adiabatic connection. This behavior results in an
unbalanced treatment of the ground and excited states.
Moreover, high-energy Rydberg excitation energies should be even
more sensitive to this effect,
as observed for transitions to the
$\P$ state.  The second variant of perturbation theory should
correct this behavior by keeping the density constant at
each order, as shown in the KS case~\cite{Gorling1995,Filippi1997}.

%Beryllium
%=====================================================================
\subsection{Beryllium atom}

When the first-order perturbation correction is applied to the ground-state
and valence-excited states of beryllium, total energies are again improved 
(although not illustrated here).
In Figure~\ref{fig:be_dacpvdz_exc_0+1}, we have plotted the zeroth- and first-order
valence excitation energies of beryllium against the range-separation parameter $\mu$.

Since valence excitation energies should be less sensitive to a poor description of the ionization
energy than Rydberg excitation energies, the first-order
correction should work better for the beryllium valence excitations than for the helium Rydberg excitations. 
However, although the singlet excitation energy of beryllium
is improved at $\mu = 0$ at first order, 
the corresponding triplet excitation energy is not improved. In fact, whereas 
the triplet excitation energy is overestimated at zeroth order, it is underestimated 
by about the same amount at first order.

As the interaction is switched on, a bump is observed
for small $\mu$ for the singlet excitation energy but not
the triplet excitation energy, which converges monotonically to its physical
limit. The convergence of the excitation energies with 
$\mu$ is improved by the first-order excitation
energies, especially in the singlet case.

%H2
%=====================================================================
\subsection{Hydrogen molecule}

In Figure~\ref{fig:h2_eq_dacpvtz_exc_0+1}, we have plotted the excitation energies of H$_2$ as a function of 
$\mu$ at the equilibrium distance $R_{\text{eq}}$ and at 3$R_{\text{eq}}$.
At the equilibrium geometry, the first-order correction works
well. At $\mu=0$, the correction is in the right direction (singlet and triplet excitation energies being
raised and lowered, respectively); for nearly all $\mu > 0$, 
the error is smaller than for the zeroth-order excitation energies. 

Unfortunately, when the bond is stretched, this is no longer the case. 
At the stretched geometry, the first excitation energy $1 ^1\Sigma_\text{g}^+ \to 1
^3\Sigma_\text{u}^+$ becomes negative for small values of $\mu$ and the error
with respect to the physical excitation energy is higher than in the
zeroth-order case.  Moreover, the ordering of the two singlet
excitation energies is incorrect at small $\mu$ and they present a
strong oscillation when the interaction is switched on. In 
this case, therefore, the zeroth-order excitation energies are better
approximations to the physical excitation energies. 

%=====================================================================
% Conclusion
\section{Conclusion}
\label{sec:conclusion}
We have considered two variants of 
perturbation theory along a range-separated adiabatic connection.
The first and simpler variant, based on the usual Rayleigh--Schr\"odinger
perturbation theory, was tested on the helium and beryllium atoms and
on the hydrogen molecule at equilibrium and stretched geometries.
Although total energies are improved to first order in the perturbation,
excitation energies are not improved since the theory does not keep the density constant
along the adiabatic connection at each
order of perturbation.  It would be interesting to
examine the evolution of the ionization potential 
to understand better the effect of this variant of the
perturbation theory on our systems of interest.

The second variant of the perturbation theory, based on 
G\"orling--Levy theory, should improve the results significantly by keeping the
ground-state density constant at each order in the perturbation~\cite{Gorling1995},
as already observed on the KS system~\cite{Filippi1997}. However, this more complicated
theory has not yet been implemented for $\mu> 0$.

An interesting alternative to perturbation theory is provided by extrapolation,
which make use of the behavior of the energies with respect to $\mu$ near the
physical system to estimate the exact energies from the energy of the
partially interacting system at a given $\mu$ and its first- or higher-order
derivatives with respect to $\mu$~\cite{Sav-JCP-11,Sav-JCP-14}. 
Work using this approach will be presented elsewhere.

\section*{Acknowledgements}
This work was supported by the Norwegian Research Council through the
CoE Centre for Theoretical and Computational Chemistry (CTCC) Grant
No.\ 179568/V30 and and through the European Research Council under
the European Union Seventh Framework Program through the Advanced
Grant ABACUS, ERC Grant Agreement No.\ 267683.

%\bibliographystyle{apsrev}

%=====================================================================
%Appendix
\appendix
%====================================================================

\section{Double adiabatic connection with a constant density}
\label{app:doubleadia}

We here present a double adiabatic connection, depending
on two parameters, that keeps the ground-state density constant.
It is the basis for the perturbation theory presented in Section~\ref{sec:secondvariant}. 
A different density-fixed double adiabatic connection 
was considered in Refs.~\onlinecite{TouGorSav-IJQC-06, Cornaton2013a}.

The Levy–Lieb universal density functional for the Coulomb
electron--electron interaction $\hat{W}_{\ee}$
is given by~\cite{Lev-PNAS-79,Lev-PRA-82,Lieb1983}
\begin{align}
  F[n] &= \min_{\Psi \to n} \bra{\Psi} \hat{T} + \hat{W}_{\ee}
  \ket{\Psi} \nonumber\\ &= \bra{\Psi[n]} \hat{T} + \hat{W}_{\ee}
  \ket{\Psi[n]},
\label{Fn}
\end{align}
We here generalize it to
the interaction $\hat{W}_{\ee}^{\lr,\mu} + \l \hat{W}_{\ee}^{\sr,\mu}$,
where $\hat{W}_{\ee}^{\lr,\mu}$ and $\hat{W}_{\ee}^{\sr,\mu}$ are
long-range and short-range electron--electron interactions,
respectively, that depend on both a range-separation parameter
$\mu$ and on a linear parameter $\l$:
\begin{align}
  F^{\mu,\l}[n] &= \min_{\Psi \to n} \bra{\Psi} \hat{T} +
  \hat{W}_{\ee}^{\lr,\mu} + \l \hat{W}_{\ee}^{\sr,\mu} \ket{\Psi}
  \nonumber\\ &= \bra{\Psi^{\mu,\l}[n]} \hat{T} +
  \hat{W}_{\ee}^{\lr,\mu} + \l \hat{W}_{\ee}^{\sr,\mu}
  \ket{\Psi^{\mu,\l}[n]}.  \nonumber\\
\label{Fmln}
\end{align}
The total universal density functional $F[n]$ is then decomposed
into $F^{\mu,\l}[n]$ and a $(\mu,\l)$-dependent
short-range Hartree--exchange--correlation density functional
$\bar{E}_{\Hxc}^{\sr,\mu,\l}[n]$,
\begin{align}
  F[n] = F^{\mu,\l}[n] + \bar{E}_{\Hxc}^{\sr,\mu,\l}[n],
\label{}
\end{align}
giving the following expression for the exact ground-state energy of
the electronic system
\begin{align}
  E_0  &= \, \min_{\Psi} \Bigl\{ \bra{\Psi} \hat{T} + \hat{V}_\text{ne} +
  \hat{W}_{\ee}^{\lr,\mu} + \l \hat{W}_{\ee}^{\sr,\mu} \ket{\Psi} \nonumber \\ & \qquad\qquad\qquad\qquad +
  \bar{E}_{\Hxc}^{\sr,\mu,\l}[n_{\Psi}]\Bigl\},
\label{E0minPsimul}
\end{align}
where the minimization is over normalized multi-determinantal wave
functions. The Euler--Lagrange equation corresponding to this
minimization is
\begin{equation}
  \hat{H}^{\mu,\l} \ket{\Psi_0^{\mu,\l}} = {\cal E}_0^{\mu,\l} \ket{\Psi_0^{\mu,\l}},
\end{equation}
where $\Psi_0^{\mu,\l}$ and ${\cal E}_0^{\mu,\l}$ are the ground-state
wave function and energy, respectively, of the Hamiltonian
\begin{equation}
  \hat{H}^{\mu,\l} = \hat{T} + \hat{V}_\text{ne} + \hat{W}_{\ee}^{\lr,\mu} +
  \l \hat{W}_{\ee}^{\sr,\mu} + \hat{\bar{V}}_{\Hxc}^{\sr,\mu,\l},
\label{Hmulapp}
\end{equation}
where 
\begin{equation}
\hat{\bar{V}}_{\Hxc}^{\sr,\mu,\l} = \int \frac{\delta \bar{E}_{\Hxc
}^{\sr,\mu,\l}[n_0] }{ \delta n(\mr)} \hat{n}(\mr) \, \mathrm d\b{r} 
\end{equation}
is the short-range Hartree--exchange--correlation potential operator, evaluated at 
the ground-state density of the physical system at $\mu$ and $\l$,
$n_0(\b{r}) = \bra{\Psi_0^{\mu,\l}} \hat{n}(\b{r})
\ket{\Psi_0^{\mu,\l}}$.
The Hamiltonian
$\hat{H}^{\mu,\l}$ thus sets up a double
adiabatic connection with a constant ground-state density. 

The range-separated ground-state DFT formalism of
Section~\ref{sec:rsdft} is recovered in the limit $\l=0$. 
To set up a perturbation theory in $\l$ about 
$0$, we rewrite $\hat{H}^{\mu,\l}$ of
Eq.~(\ref{Hmulapp}) as the sum of the noninteracting Hamiltonian 
$\hat{H}^{\lr,\mu} = \hat{H}^{\mu,\l=0}$ and a perturbation
operator. For this purpose, the Hartree--correlation--exchange functional
is written as
\begin{equation}
  \bar{E}_{\Hxc}^{\sr,\mu,\l}[n] = \bar{E}_{\Hxc}^{\sr,\mu,\l=0}[n]
  - E_{\Hxc}^{\sr,\mu,\l}[n],
\label{EHxcsrmulapp}
\end{equation}
which defines the new functional $E_{\Hxc}^{\sr,\mu,\l}[n]$. The
Hamiltonian can now be rewritten as
\begin{equation}
  \hat{H}^{\mu,\l} = \hat{H}^{\lr,\mu} + \l \hat{W}_{\ee}^{\sr,\mu} -
  \hat{V}_{\Hxc}^{\sr,\mu,\l},
\label{Hmulapp2}
\end{equation}
where 
\begin{equation}
  \hat{V}_{\Hxc}^{\sr,\mu,\l}= \int \frac{\delta
E_{\Hxc}^{\sr,\mu,\l}[n_0]}{\delta n(\mr)} \, \hat{n}(\mr) \, \mathrm d\b{r}
\end{equation} is the
short-range Hartree--exchange--correlation potential operator associated
with $E_{\Hxc}^{\sr,\mu,\l}[n]$.

The dependence on $\l$ of $E_{\Hxc}^{\sr,\mu,\l}[n]$
can be made more explicit. It is easy to show that
\begin{align}
    E_{\Hxc}^{\sr,\mu,\l}[n] &=
     \bra{\Psi^{\mu,\l}[n]} \hat{T} +
    \hat{W}_{\ee}^{\lr,\mu} + \l \hat{W}_{\ee}^{\sr,\mu}
    \ket{\Psi^{\mu,\l}[n]} \nonumber \\ &- \bra{\Psi^{\mu,\l=0}[n]}
  \hat{T} + \hat{W}_{\ee}^{\lr,\mu} \ket{\Psi^{\mu,\l=0}[n]},
\end{align}
which leads to the following decomposition
\begin{equation}
  E_{\Hxc}^{\sr,\mu,\l}[n] = \l E_{\H \x,\md}^{\sr,\mu}[n] + E_{\tc,\md}^{\sr,\mu,\l}[n],
\label{Ehxcmddecomp}
\end{equation}
where
\begin{equation}
E_{\H \x,\md}^{\sr,\mu}[n] = \bra{\Psi^{\mu,\l=0}[n]}
\hat{W}_{\ee}^{\sr,\mu} \ket{\Psi^{\mu,\l=0}[n]}
\end{equation} 
is a
multi-determinantal (md) generalization of the usual short-range
Hartree--exchange
functional~\cite{TouGorSav-TCA-05,GorSav-IJQC-09a,Stoyanova2013}. Using
the variational property of the wave function $\Psi^{\mu,\l}[n]$, and
for non-degenerate wave functions $\Psi^{\mu,\l=0}[n]$, the expansion
of $E_{\tc,\md}^{\sr,\mu,\l}[n]$ in $\l$ about $0$ starts at second
order: 
\begin{equation}
E_{\tc,\md}^{\sr,\mu,\l}[n] = \l^2 E_{\tc,\md}^{\sr,\mu,(2)}[n] + \cdots, 
\end{equation}
as in standard GL perturbation
theory~\cite{GorLev-PRB-93,Gorling1995}. The Hamiltonian
of Eq.~(\ref{Hmulapp2}) can now be rewritten as
\begin{equation}
  \hat{H}^{\mu,\l} = \hat{H}^{\lr,\mu} + \l \hat{W}^{\sr,\mu} -
  \hat{V}_{\tc,\md}^{\sr,\mu,\l},
\label{Hmulapp3}
\end{equation}
where the perturbation operator $\hat{W}^{\sr,\mu} =
\hat{W}_{\ee}^{\sr,\mu} - \hat{V}_{\H \x,\md}^{\sr,\mu}$ and
\begin{equation}
\hat{V}_{\H \x,\md}^{\sr,\mu}= \int \frac{\delta E_{\H \x,\md}^{\sr,\mu}[n_0]}{\delta n(\mr)} \hat{n}(\mr) \, \mathrm d\b{r} 
\end{equation}
has been introduced to collect
all the linear terms in $\l$, the remaining perturbation operator
\begin{equation}
\hat{V}_{\tc,\md}^{\sr,\mu,\l} = \int \frac{\delta
E_{\tc,\md}^{\sr,\mu,\l}[n_0] }{ \delta n(\mr)} \hat{n}(\mr) \, \mathrm d\b{r}
\end{equation}
containing all higher-order terms in $\l$.

\

\end{document}